\documentclass{article}
\usepackage[T1]{fontenc}
\usepackage{graphicx} 
\usepackage{bbm}
\usepackage{placeins}
\usepackage{hyperref}
\usepackage{appendix}
\usepackage{amsmath}
\usepackage[symbol]{footmisc}
\usepackage{tablefootnote}
\usepackage[nonatbib, final]{neurips_2024}

\title{%
  \begin{tabular}{@{}c@{\hspace{1em}}l@{}}
    \includegraphics[height=3em]{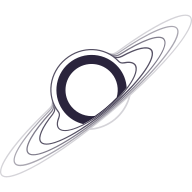} &
    \parbox[b]{0.8\textwidth}{\raggedright\bfseries
      CustomIR: Unsupervised Fine-Tuning of Dense Embeddings for Known Document Corpora
    }
  \end{tabular}%
}

\author{Nathan Paull \\
  \texttt{Valkyrie Andromeda}
}

\date{\today}

\begin{document}

\maketitle
\begin{abstract}
Dense embedding models have become critical for modern information retrieval, particularly in RAG pipelines, but their performance often degrades when applied to specialized corpora outside their pre-training distribution. To address this we introduce \textbf{CustomIR}, a framework for unsupervised adaptation of pre-trained language embedding models to domain-specific corpora using synthetically generated query–document pairs. CustomIR leverages large language models (LLMs) to create diverse queries grounded in a known target corpus, paired with LLM-verified hard negatives, eliminating the need for costly human annotation. Experiments on enterprise email and messaging datasets show that CustomIR consistently improves retrieval effectiveness with small models gaining up to 2.3 points in Recall@10. This performance increase allows these small models to rival the performance of much larger alternatives, allowing for cheaper RAG deployments. These results highlight that targeted synthetic fine-tuning offers a scalable and cost-efficient strategy for increasing domain-specific performance.
\end{abstract}

\section*{Introduction}

Dense embedding models have become the backbone of modern information retrieval, largely due to training on massive, heterogeneous datasets that capture broad linguistic patterns \cite{beir, mteb, mmteb}. In practice, these models are designed to handle many domains simultaneously, leading to a strategy of scaling both model and dataset in order to improve benchmark performance \cite{scaling,embedding_limits}. While effective for broad coverage, this approach is computationally expensive and often intractable for real-world implementation of dense embeddings, even exceeding the domain needs of many applications where users operate within a single domain.

We argue that the challenges faced by dense embeddings in specialized domains—such as legal, biomedical, or enterprise texts, stem less from insufficient coverage and more from an adaptation problem: inference data differs greatly from the training corpus, leading to data distribution imbalances or even out-of-distribution (OOD) effects. Rather than attempting to scale models and datasets to encompass all possible domains, it is more efficient to adapt and specialize smaller models to the target domain. To this end, we introduce \textbf{CustomIR}, a synthetic data generation framework that adapts pre-trained embeddings to a target corpus in an unsupervised manner. CustomIR generates domain-specific query–document pairs via large language models (LLMs), enabling fine-tuning without human annotation. This approach captures domain-specific terminology and end-user retrieval patterns, yielding substantial gains in specialized retrieval tasks while requiring only a fraction of the resources needed for scaling-based methods.

\section*{Research Priors}

\subsection*{Natural Language Embedding}
Natural language embeddings are a central component of modern information retrieval and representation learning systems. Early approaches were dominated by predictive word embedding models such as Word2Vec \cite{word2vec} and GloVe \cite{glove}, which demonstrated the ability to capture semantic similarity in dense vector spaces. These static embeddings were later superseded by contextualized representations introduced by models such as ELMo \cite{elmo}, which leveraged deep recurrent architectures to generate embeddings sensitive to linguistic context. 

The advent of transformer-based models, most prominently BERT \cite{bert}, marked a significant shift by enabling bidirectional contextual understanding at scale. Variants and extensions of BERT, such as RoBERTa \cite{roberta} and ALBERT \cite{albert}, improved pretraining efficiency and embedding quality, making them widely adopted for downstream fine-tuning tasks. In parallel, sentence-level representation models such as Sentence-BERT (SBERT) \cite{sbert} specialized BERT-style encoders for semantic similarity and retrieval tasks, producing embeddings optimized for pairwise comparison. 

Recent work has further advanced embedding models through large-scale contrastive learning \cite{simcse}, domain-specific pretraining \cite{adapt}, and multilingual extensions \cite{agnostic_bert}, expanding their applicability to diverse tasks and languages. Moreover, models such as OpenAI’s text-embedding series and instruction-tuned encoders have demonstrated that fine-tuning embeddings with task-oriented supervision can yield substantial improvements in retrieval performance. 

\subsection*{LLM-Generated Synthetic Data}

Synthetic data generation with large language models has emerged as a promising alternative to traditional large-scale data collection. Rather than scaling datasets indiscriminately, this approach focuses on generating high-quality task-relevant examples to address model deficiencies. Prior work has shown that carefully constructed synthetic data can substantially improve performance in reasoning, language understanding, and retrieval tasks.

For example, DeepSeek R1 demonstrates how LLM-generated synthetic reasoning chains and mathematical problems can improve logical inference and problem-solving abilities that are underrepresented in naturally occurring corpora \cite{r1}. The use of this synthetic reasoning data in the fine-tuning stage improves emergent thinking capabilities in the R1 model. Similarly, the Phi models achieve SOTA performance on small parameter language models through the use of high-quality synthetic, textbook-style data \cite{phi1,phi1_5,phi3,phi4}. 

Synthetic data has emerged as a powerful tool for improving dense embeddings in information retrieval. For example, \textbf{Qwen3-Embed} leverages diverse query–document pairs generated from multiple query personas, yielding strong gains on MTEB benchmarks\cite{qwen3}. More broadly, recent studies show that synthetic data can rival or even surpass human-annotated pairs in training effectiveness, making it a scalable alternative for embedding model development\cite{synthetic_data}.

One limitation of synthetic data is the propensity for LLMs to output stylistically uniform text, leading to low output diversity and, therefore, low quality synthetic data. A particularly important development to address this issue is the use of persona-driven generation. By varying the "voice" or perspective of the LLM, models can be trained on a richer distribution of linguistic intents\cite{scalingsyntheticdatacreation}. This technique has been shown to reduce bias toward a single query style and to capture edge cases that otherwise remain unrepresented\cite{synthetic_survey}. Such strategies are especially valuable for domain adaptation, where the goal is to capture the full spectrum of domain-specific patterns without requiring expensive human annotation or data capture.

\section*{Methodology}

\subsection*{Dataset Selection and Motivation}
For our experiments, we use two enterprise communication datasets: the Enron email subset of the ConcurrentQA dataset\cite{cqa} and internal Valkyrie Slack messages from Q1 2025. Both corpora closely resemble enterprise-style internal communications at different scales. We exclude the Wikipedia portion of ConcurrentQA because baseline models achieved near-ceiling performance on these documents, likely due to pre-training exposure (see Appendix \hyperref[sec:appendix_b]{B}). It is worth noting that the Enron dataset is already pre-processed to work directly with most off-the-shelf embedding models. For the Slack dataset we treat each message as its own document for simplicity. 

\FloatBarrier
\begin{table}[htb]
    \centering
    \begin{tabular}{lcc}
        \hline
        Dataset & Documents & Avg. Length (char) \\
        \hline
        Enron Email & 10,077 & 1069.7 \\
        Slack Messages & 1,678 & 784.0 \\
        \hline
    \end{tabular}
    \caption{Dataset statistics for the enterprise communication corpora used in our experiments.}
    \label{tab:dataset_stats}
\end{table}
\FloatBarrier

Using both email and Slack data allows us to study retrieval performance across different enterprise communication modalities without introducing significant pre-processing overhead, while demonstrating the generalizability of our approach across similar but distinct workplace domains.

\subsection*{Model Selection}

Table~\ref{tab:model_specs} summarizes the embedding models evaluated in our study. We selected BGE-M3\cite{bge} and Qwen3-Embed-Sm\cite{qwen3} as primary baselines due to their strong general-domain performance, architectural differences, and high throughput without quantization. These smaller models provide an effective testbed for zero-shot performance on enterprise-style corpora and serve as efficient training targets for CustomIR fine-tuning.

\FloatBarrier
\begin{table}[htb]
    \centering
    \begin{tabular}{lcccc}
        \hline
        Model & Params & Embed. Dim. & Pooling & Prompting \\
        \hline
        BGE-M3 & 0.6B & 1024 & Mean & None \\
        Qwen3-Embed-Sm & 0.6B & 1024 & Last Token & Query Only \\
        Qwen3-Embed-Md & 4.0B & 2560 & Last Token & Query Only \\
        Qwen3-Embed-Lg & 8.0B & 4096 & Last Token & Query Only \\
        \hline
    \end{tabular}
    \caption{Model specifications for the embedding models used in our experiments.}
    \label{tab:model_specs}
\end{table}
\FloatBarrier

To contextualize the benefits of domain adaptation, we also include Qwen3-Embed-Md and Qwen3-Embed-Lg, which represent larger and more computationally expensive alternatives. This allows us to assess whether fine-tuning smaller models with CustomIR can rival the performance of substantially larger models, highlighting the trade-off between scale and targeted adaptation. Finally, because Qwen3-Embed supports prompted embeddings, we provide a detailed comparison of task prompts in Appendix \hyperref[sec:appendix_c]{C}, where we show how different retrieval instructions affect downstream performance. For all evaluations of Qwen3-Embed we use the task prompt, "Retrieval relevant passage for the given query" pre-pended to each query.

\subsection*{Synthetic Dataset Generation}

\FloatBarrier
\begin{figure}[htb]
    \centering
    \includegraphics[width=1\linewidth]{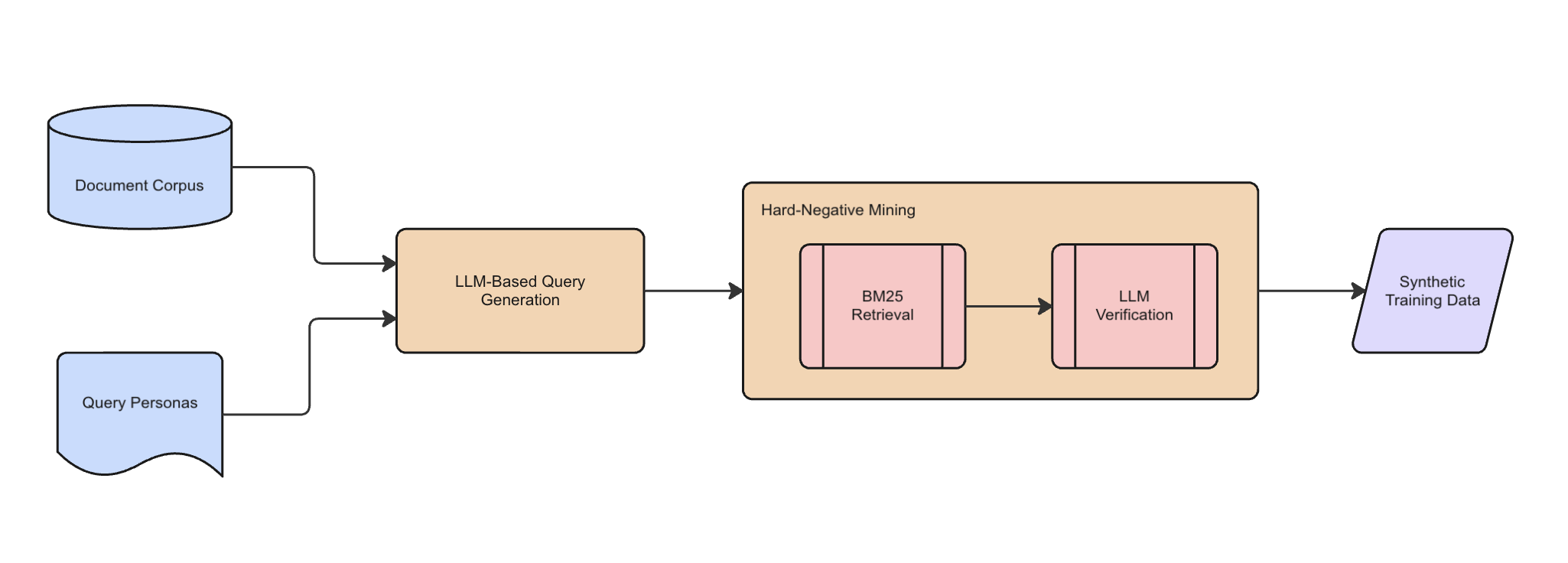}
    \caption{The CustomIR Synthetic Dataset Generation Pipeline.}
    \label{fig:data_pipe}
\end{figure}
\FloatBarrier

In CustomIR, synthetic query-document pairs are generated for each retrieval chunk using an LLM, for our experiments we use GPT-4o\cite{gpt4o}. To capture diverse retrieval behaviors and prevent response collapse, we employ multiple query personas, such as keyword-based, open-ended, and task-oriented queries. These pairs of synthetic queries and real document chunks form positive examples for fine-tuning. To support contrastive supervision, we augment each query with \textbf{k} hard negatives mined using BM25\cite{bm25} on the complete document corpus. Each negative is verified by an LLM to remove any false negatives and ensure a strong and accurate contrastive signal. This approach produces a high-quality, domain-specific synthetic dataset that requires no human annotation.

\subsection*{Training and Evaluation}

We fine-tune dense embedding models using an optimized contrastive loss, specifically the Qwen3-Embed variant of InfoNCE \cite{qwen3,infonce}. Given a batch of $N$ training instances, the embedding loss is defined as:

\begin{equation}
\mathcal{L}_{\text{embedding}} = -\frac{1}{N} \sum_{i=1}^{N} 
\log \frac{\exp\left(s(q_i, d_i^{+}) / \tau \right)}{Z_i},
\end{equation}

where $s(\cdot, \cdot)$ is cosine similarity, $\tau$ is a temperature parameter, and $Z_i$ is the partition function aggregating positive and negative similarities:

\begin{equation}
\begin{aligned}
Z_i = & \ \exp\left(s(q_i, d_i^{+}) / \tau\right) 
+ \sum_{k=1}^{K} m_{ik} \exp\left(s(q_i, d_{i,k}^{-}) / \tau\right) \\
& + \sum_{j \neq i} m_{ij} \exp\left(s(q_i, q_j) / \tau\right) 
+ \sum_{j \neq i} m_{ij} \exp\left(s(d_i^{+}, d_j) / \tau\right) \\
& + \sum_{j \neq i} m_{ij} \exp\left(s(q_i, d_j) / \tau\right).
\end{aligned}
\end{equation}

Here $d_i^{+}$ denotes the positive document for query $q_i$, $d_{i,k}^{-}$ are $K$ hard negatives, and $m_{ij}$ is a masking factor to reduce the influence of false negatives, defined as:

\begin{equation}
m_{ij} =
\begin{cases}
0 & \text{if } s_{ij} > s(q_i, d_i^{+}) + 0.1 \ \text{or} \ d_j = d_i^{+}, \\
1 & \text{otherwise}.
\end{cases}
\end{equation}

Training batches contain one positive and multiple verified negatives per query. We optimize with AdamW \cite{adamw}, using linear learning rate warmup and decay, and apply early stopping based on validation loss. For more information on training details see Appendix \hyperref[sec:appendix_a]{A}.

To create an evaluation dataset, we enforce a strict query-level split on our full synthetic dataset so that each query and its associated relevance judgments appear exclusively in either training or evaluation. This allows for the entire corpus to exist in both the training and evaluation datasets while maintaining query and query relevance uniqueness between the two datasets. Performance is measured using \textbf{Recall@k} to quantify retrieval coverage and \textbf{nDCG@k} to assess ranking quality, providing a comprehensive view of model performance on enterprise-style corpora.

\newpage
\section*{Results}

\FloatBarrier
\begin{table}[htb]
    \centering
    \begin{tabular}{lcccccc}
        \hline
        Model & R@10 & R@50 & R@100 & nDCG@10 & nDCG@50 \\
        \hline
        BGE-M3 & 0.5411 & 0.7131 & 0.7896 & 0.6266 & 0.6665 \\
        BGE-M3 + CustomIR & 0.5570 & 0.7291 & 0.7973 & 0.6472 & 0.6887 \\
        Qwen3-Embed-Sm & 0.5504 & 0.7209 & 0.7950 & 0.6480 & 0.6834 \\
        Qwen3-Embed-Sm + CustomIR & \textbf{0.5664} & \textbf{0.7351} & \textbf{0.7988} & \textbf{0.6597} & \textbf{0.6974} \\
        \hline
        Qwen3-Embed-Md & 0.5677 & 0.7492 & 0.8217 & 0.6695 & 0.7086 \\
        Qwen3-Embed-Lg & 0.5702 & 0.7541 & 0.8241 & 0.6674 & 0.7083 \\
        \hline
    \end{tabular}
    \caption{Retrieval performance on the Enron email dataset. Recall@k (R@k) and nDCG@k metrics are reported. "+ CustomIR" indicates fine-tuning with synthetic domain-adaptation.}
    \label{tab:enron_results}
\end{table}
\FloatBarrier

\FloatBarrier
\begin{table}[htb]
    \centering
    \begin{tabular}{lcccccc}
        \hline
        Model & R@10 & R@50 & R@100 & nDCG@10 & nDCG@50 \\
        \hline
        BGE-M3 & 0.6924 & 0.8446 & 0.8902 & 0.6734 & 0.7285 \\
        BGE-M3 + CustomIR & 0.7034 & 0.8541 & 0.9039 & 0.6872 & 0.7404 \\
        Qwen3-Embed-Sm & 0.7098 & 0.8491 & 0.8891 & 0.6848 & 0.7335 \\
        Qwen3-Embed-Sm + CustomIR & \textbf{0.7305} & \textbf{0.8772} & \textbf{0.9204} & \textbf{0.7114} & \textbf{0.7648} \\
        \hline
        Qwen3-Embed-Md & 0.7243 & 0.8752 & 0.9132 & 0.7274 & 0.7809 \\
        Qwen3-Embed-Lg & 0.7415 & 0.8787 & 0.9150 & 0.7279 & 0.7777 \\
        \hline
    \end{tabular}
    \caption{Retrieval performance on the Slack messages dataset. Recall@k (R@k) and nDCG@k metrics are reported. "+ CustomIR" indicates fine-tuning with synthetic domain-adaptation.}
    \label{tab:slack_results}
\end{table}
\FloatBarrier

Tables~\ref{tab:enron_results} and~\ref{tab:slack_results} show that CustomIR consistently improves retrieval performance across both enterprise datasets and embedding models. On Enron, CustomIR yields modest gains for BGE-M3 but larger gains for Qwen3-Embed-Sm, improving Recall@10 by +1.6 points and nDCG@10 by +1.2 points over its baseline. On Slack, the improvements are similar with BGE-M3 gains +1.1 R@10 and +1.4 nDCG@10, while Qwen3-Embed-Sm achieves +2.1 R@10 and +2.6 nDCG@10 after CustomIR fine-tuning.

A key observation is that CustomIR-adapted Qwen3-Embed-Sm narrows the gap with the much larger Qwen3-Embed-Md and Lg models. On Enron, the fine-tuned small model approaches the performance of Qwen3-Embed-Md, while on Slack it surpasses it in Recall@10 (0.7305 vs.\ 0.7243) and is within 1--2 points of the Lg variant in both recall and nDCG. This indicates that targeted domain adaptation of smaller models can rival, and in some cases match, the performance of larger, more computationally expensive alternatives.

Taken together, these results demonstrate that synthetic domain adaptation via CustomIR is an efficient and effective alternative to traditional dataset and capacity scaling. By closing the gap between compact and large models without requiring human annotation or significant compute overhead, CustomIR provides a practical solution for enterprise retrieval adaptation.

\subsection*{Ablation Study}

To better understand the contribution of each component of our domain adaptation framework, we conduct an ablation over three training regimes, Baseline (no adaptation), Synthetic (fine-tuning with synthetic queries and BM25-mined negatives), and CustomIR (our full method with LLM-based filtering of false negatives).

\FloatBarrier
\begin{table}[htb]
    \centering
    \begin{tabular}{lcc}
        \hline
        & Slack & Enron \\
        \hline
        Mined negatives per query & 10 & 25 \\
        LLM-verified false negative rate   & 19.68\% & 31.69\% \\
        \hline
    \end{tabular}
    \caption{Hard negatives mined with BM25 exhibit a substantial false negative rate, motivating the use of LLM-based verification.}
    \label{tab:false_negatives}
\end{table}
\FloatBarrier

To motivate the ablation, we first examine the quality of BM25-mined negatives across domains. Table~\ref{tab:false_negatives} shows that the BM25 retrievals for Slack and Enron contain a large number of positive documents as judged by an LLM. This high noise reduces the faithfulness of a synthetic-only training set, as mislabeled positives are treated as negatives during contrastive learning.

\FloatBarrier
\begin{table}[htb]
\centering
    \begin{tabular}{lccccc}
        \hline
        Model & R@10 & R@50 & R@100 & nDCG@10 & nDCG@50 \\
        \hline
        Qwen3-Embed-Sm (Baseline) & 0.5504 & 0.7209 & 0.7950 & 0.6480 & 0.6834 \\
        + Synthetic Negatives & 0.5531 & 0.7135 & 0.7764 & 0.6519 & 0.6849 \\
        + CustomIR (LLM-verified) & \textbf{0.5664} & \textbf{0.7351} & \textbf{0.7988} & \textbf{0.6597} & \textbf{0.6974} \\
        \hline
    \end{tabular}
    \caption{Ablation on the Enron dataset. Synthetic negatives alone yield only marginal improvements, while CustomIR consistently outperforms both baseline and synthetic-only training.}
    \label{tab:enron_ablation}
\end{table}
\FloatBarrier

\FloatBarrier
\begin{table}[htb]
    \centering
    \begin{tabular}{lccccc}
        \hline
        Model & R@10 & R@50 & R@100 & nDCG@10 & nDCG@50 \\
        \hline
        Qwen3-Embed-Sm (Baseline) & 0.7098 & 0.8491 & 0.8891 & 0.6848 & 0.7335 \\
        + Synthetic Negatives & 0.7207 & 0.8633 & 0.8991 & \textbf{0.7168} & \textbf{0.7680} \\
        + CustomIR (LLM-verified) & \textbf{0.7305} & \textbf{0.8772} & \textbf{0.9204} & 0.7114 & 0.7648 \\
        \hline
    \end{tabular}
    \caption{Ablation on the Slack dataset. Starting from the baseline encoder, adding synthetic queries and BM25 negatives improves retrieval, while the full CustomIR pipeline yields the best results.}
    \label{tab:slack_ablation}
\end{table}
\FloatBarrier

On Slack, LLM filtering consistently improves recall over synthetic negatives alone while showing aslight loss in performance in nDCG. On the more challenging Enron corpus, the effect is even clearer, adding LLM verification produces consistent gains across all metrics. These results confirm that the LLM filtering step is essential to adapting IR performance with mined negatives, particularly in noisier domains where many positive documents are more prevalent.

\section*{Conclusion}

In this work, we introduced \textbf{CustomIR}, a synthetic data generation framework that adapts pre-trained dense embedding models to specialized domains in an unsupervised manner. By generating diverse query–document pairs tailored to a target corpus and augmenting them with verified hard negatives, CustomIR enables efficient fine-tuning and enhances retrieval effectiveness across both retrieval coverage and ranking quality metrics.

Our experiments on enterprise communication datasets show that CustomIR consistently improves retrieval performance across multiple embedding models. In particular, the CustomIR-adapted Qwen3-Embed-Sm rivals, and in some cases surpasses, the performance of much larger Qwen3-Embed-Md variant, achieving comparable recall and ranking quality at a fraction of the computational cost. These results highlight that targeted adaptation of smaller models can close the performance gap with larger ones, providing a more practical and resource-efficient alternative to indiscriminate model and dataset scaling.

\textbf{Limitations: }Our evaluation is limited to two enterprise communication domains (emails and Slack messages), as such these results may not generalize to other specialized fields such as medical or legal texts. Our binary verification approach may also lose nuanced similarity information captured by continuous scoring methods. Additionally, the quality of synthetic data depends on the capabilities and potential biases of the LLM and query personas, which could systematically affect retrieval performance. Finally, our synthetic benchmark assumes that the query distribution remains stable between training and evaluation, while real-world use cases will inevitably exhibit distribution drift. Addressing this distributional mismatch remains a key challenge for future work.

\nocite{sbert, bert,qwen, e5, mteb, mmteb, facebook_search, scalingsyntheticdatacreation, scaling, synthetic_data, followir}

\newpage

{\small
\bibliographystyle{ieee_fullname}
\bibliography{refs}
}

\newpage\appendix

\section*{Appendix A: Training Details}\label{sec:appendix_a}

We fine-tuned both Qwen3-Embed and BGE-M3 using similar hyperparameter settings to ensure a fair comparison. Table~\ref{tab:training_details} summarizes the key training configuration for each dataset.

\begin{table}[htb]
    \centering
    \begin{tabular}{lcc}
        \hline
        \textbf{Configuration} & \textbf{Enron Emails} & \textbf{Slack Messages} \\
        \hline
        Loss Function & QwenInfoNCE & QwenInfoNCE \\
        Similarity & Cosine & Cosine \\
        Temperature & 0.07 & 0.07 \\
        Reduction & Mean & Mean \\
        Max Sequence Length & 1024 & 1024 \\
        Global Train Batch Size & 64 & 16 \\
        Hard Negatives per Sample & 20 & 8 \\
        Optimizer & AdamW & AdamW \\
        Learning Rate & $1\times 10^{-5}$ & $1\times 10^{-6}$ \\
        Weight Decay & 0.01 & 0.01 \\
        Scheduler & Linear Decay with Warmup & Linear Decay with Warmup \\
        Warmup Proportion & 0.06 & 0.06 \\
        Final LR Factor ($\alpha_f$) & 0.02 & 0.02 \\
        Seed & 42 & 42 \\
        Precision & AMP BF16 & AMP BF16 \\
        \hline
    \end{tabular}
    \caption{Summary of training configurations for Qwen3-Embed and BGE-M3 fine-tuning on the Enron email dataset.}
    \label{tab:training_details}
\end{table}

All other parameters followed standard defaults. All training was conducted on A100 GPUs using mixed precision (AMP BF16). Validation loss was computed after each epoch to support early stopping.

\section*{Appendix B: Rationale for Excluding Wikipedia Data from Benchmark}\label{sec:appendix_b}

While the ConcurrentQA dataset contains both Wikipedia article and Enron email chunks, our benchmark focuses exclusively on the Enron email subset. Table~\ref{tab:enron_results} compares the baseline retrieval performance of BGE-M3 and Qwen3-Embed-Sm on Enron emails and Wikipedia-derived chunks.

\FloatBarrier
\begin{table}[htb]
    \centering
    \begin{tabular}{lccccccc}
        \hline
        Model & Split & R@10 & R@50 & R@100 & nDCG@10 & nDCG@50 \\
        \hline
        BGE-M3 & Enron & 0.5211 & 0.6871 & 0.7642 & 0.6017 & 0.6395 \\
         & Wiki & 0.9191 & 0.9341 & 0.9386, & 0.9271 & 0.9312 \\
        \hline
        Qwen3-Embed-Sm & Enron & 0.5477 & 0.7194 & 0.7913 & 0.6459 & 0.6819 \\
         & Wiki & 0.9214 & 0.9390 & 0.9432 & 0.9286 & 0.9334\\
        \hline
    \end{tabular}
    \caption{Comparison of out-of-the-box performance for Enron Emails and Wikipedia subsets of the ConcurrentQA dataset.}
    \label{tab:wiki_comp}
\end{table}
\FloatBarrier

These results demonstrate that both models achieve near-ceiling retrieval performance on Wikipedia chunks. This is consistent with the idea that Wikipedia was most likely part of the pre-training corpora for both models, a fact confirmed for BGE-M3 \cite{bge,xlm_roberta}. As a result, performance on Wikipedia does not provide a meaningful assessment of model adaptation or domain-specific retrieval capabilities.

In contrast, the Enron subset represents a distinct, specialized domain that differs substantially from the models' pre-training data, simulating internal private data. As shown in Table~\ref{tab:wiki_comp}, performance drops significantly for both models when retrieving from Enron emails, highlighting the challenges posed by domain shift. By focusing on the Enron emails, our benchmark better captures realistic OOD retrieval scenarios and the effectiveness of domain-adaptive techniques, whereas Wikipedia performance would obscure these effects due to strong pre-training overlap.

\section*{Appendix C: Prompt Selection for Qwen3-Embed}\label{sec:appendix_c}

Qwen3-Embed supports \textit{prompted embeddings}, where different task-specific prompts are pre-pended to queries during training and inference. Each task is expressed in a structured format of the form: \texttt{"Instruct: \{task\}\textbackslash nQuery:"}, where the \texttt{\{task\}} placeholder specifies the retrieval task instruction. Since prompt choice directly influences retrieval effectiveness, we conducted controlled experiments comparing multiple task formulations on both the Slack and Enron corpora.

\subsection*{Prompt Definitions}

We evaluated the following candidate task instructions from the Qwen3-Embed repository\cite{qwen3}:

\begin{itemize}
    \item[P1:] \texttt{"Given a web search query, retrieve relevant passages that answer the query"}
    \item[P2:] \texttt{"Given a question, retrieve passages that answer the question"}
    \item[P3:] \texttt{"Retrieval relevant passage for the given query"}
    \item[P4:] \texttt{"Given a question about coding, retrieval code or passage that can solve user's question"}
\end{itemize}

\subsection*{Quantitative Results}

We report Recall@k and nDCG@k for each prompt across two domains: Slack communications and Enron emails. 

\FloatBarrier
\begin{table}[htb]
\centering
\small
\begin{tabular}{lcccc}
\hline
\textbf{Prompt ID} & \textbf{recall@10} & \textbf{recall@50} & \textbf{recall@100} & \textbf{nDCG@10} \\
\hline
\multicolumn{5}{c}{\textbf{Slack Dataset}} \\
\hline
P1 & 0.699 & 0.845 & 0.887 & 0.682 \\
P2 & 0.570 & 0.715 & 0.787 & 0.546 \\
P3 & 0.710 & 0.849 & 0.889 & 0.685 \\
P4 & 0.702 & 0.832 & 0.882 & 0.678 \\
\hline
\multicolumn{5}{c}{\textbf{Enron Dataset}} \\
\hline
P1 & 0.548 & 0.719 & 0.791 & 0.646 \\
P2 & 0.514 & 0.681 & 0.747 & 0.589 \\
P3 & 0.550 & 0.721 & 0.795 & 0.648 \\
P4 & 0.542 & 0.716 & 0.789 & 0.636 \\
\hline
\end{tabular}
\caption{Comparison of Qwen3-Embed prompt formulations across Slack and Enron datasets. Best results per column highlighted in bold.}
\label{tab:qwen3-prompt-results}
\end{table}
\FloatBarrier

\subsection*{Discussion}

Across both datasets, P3 (\texttt{"Retrieval relevant passage for the given query"}) slightly outperforms other task prompts, particularly in small retrieval domains, Recall@10 and nDCG@10. Consequently, P3 was selected as the default prompt in our training and evaluation pipeline.

\end{document}